\documentclass[non-peer-review]{fa2026}

\usepackage{xfrac}

\newcommand*{\tran}{^\mathrm{T}}
\newcommand*{\norm}[1]{\left\lVert#1\right\rVert}
\newcommand*{\ceil}[1]{\lceil#1\rceil}
\newcommand*{\floor}[1]{\lfloor#1\rfloor}
\makeatletter
\newcommand*{\rom}[1]{\expandafter\@slowromancap\romannumeral #1@}
\makeatother
\DeclareMathOperator{\diag}{diag}
\DeclareMathOperator{\sign}{sign}
\newcommand*{\FvK}{F{\"o}ppl--von K{\'a}rm{\'a}n}
\newcommand*{\mybox}[1]{\raisebox{0pt}[0pt][0pt]{#1}}

\renewcommand{\epsilon}{\varepsilon}
\renewcommand{\theta}{\vartheta}
\renewcommand{\kappa}{\varkappa}
\renewcommand{\phi}{\varphi}

\usepackage{siunitx}
\AtEveryBibitem{\clearfield{doi}}

\title{Explicit and Stable Pseudospectral Time-Domain Method for the \FvK{} Equations}

\author[1]{Victor Zheleznov}
\author[2]{Stefan Bilbao}
\correspondingauthor{v.zheleznov@ed.ac.uk}{Victor Zheleznov et al.}

\affil[1]{Acoustics and Audio Group, University of Edinburgh, Edinburgh, UK}
\affil[2]{STMS (UMR9912), IRCAM, CNRS, Sorbonne Universit{\'e}, Paris, France}

\begin{document}
\maketitle

\begin{abstract}
Modal synthesis is a widely-used technique for simulation of musical instrument dynamics. In the linear case, a modal decomposition leads to an uncoupled system of damped and forced harmonic oscillators which can be efficiently solved by standard time-stepping methods. However, extensions to nonlinear problems are challenging due to the presence of products of modal expansions in the governing equations. In the case of the \FvK{} plate, the nonlinear coupling between the modes is described by a fourth-order tensor and is prohibitively expensive to evaluate in the modal domain. In this work, we propose a pseudospectral method in which the products are evaluated on a grid in the spatial domain while spatial derivatives are computed exactly in the modal domain. Discrete sine and cosine transforms between the modal and spatial domains are used to impose simply supported boundary conditions for the plate. Finally, we prove non-negativity of the nonlinear potential energy of the system and employ a scalar auxiliary variable technique for explicit and stable time integration in the modal domain. As a result, we reduce the computational cost of modal synthesis while preserving its advantages like a precise control over the simulated frequency range. Sound examples are presented.
\end{abstract}


\section{Introduction}

High-amplitude vibration of a thin plate is perhaps the most dramatic example of nonlinear behaviour in musical acoustics, leading to characteristic pitch glides and crashes in percussion instruments. The \FvK{} equations are commonly used to reproduce these perceptually-important effects in audio rate simulations~\cite{BilbaoNSS}. One of the techniques to numerically solve such equations is known as modal synthesis~\cite{Ducceschi2015}. This approach falls into a category of Galerkin spectral methods where the governing equations are made orthogonal to basis functions of a finite series that approximate the solution~\cite{BoydSpectral}. In the case of the \FvK{} equations, this leads to a computationally-expensive model~\cite{Ducceschi2015} which requires a GPU for real-time applications~\cite{Diaz2025}.

In this work, we employ a pseudospectral method where the governing equations are satisfied for a grid of collocation points in the spatial domain~\cite{BoydSpectral}. Such methods are well-suited for nonlinear problems since they give rise to vector products rather than functional series that require determination of expansion coefficients~\cite{FornbergPS}. We decompose unknown functions using Fourier sine and cosine series to satisfy simply supported boundary conditions for a rectangular plate~\cite{Wise2021}. For time integration, we rewrite the governing equations in the modal domain to ensure exact discretisation of the linear part of the problem~\cite{BilbaoNSS} and employ a scalar auxiliary variable technique~\cite{Risse2025}. This results in an explicit and stable numerical solver for the \FvK{} equations in contrast to an implicit and iterative scheme previously used by Kirby and Yosibash~\cite{Kirby2004}. Compared to a modal approach developed by Russo~et~al.\@~\cite{Russo2025} in the related context of nonlinear string vibration, spatial derivatives are computed exactly in the modal domain rather than approximated by finite differences in the spatial domain.

The paper is organised as follows. The continuous model of a thin plate is given in Section~\ref{sec: cont_model}. In Section~\ref{sec: ps} modal equations using the pseudospectral method are derived, and then discretised in time using an explicit and stable numerical solver in Section~\ref{sec: solver}. Numerical results are presented in Section~\ref{sec: results}. Sound examples are available on the accompanying page\footnote{\url{https://victorzheleznov.github.io/fa2026}}.

\section{Continuous Model}\label{sec: cont_model}

The \FvK{} equations in a scaled form over a unit area region $\mathcal{D}$ are given as follows~\cite{BilbaoNSS}:
\begin{subequations}\label{eq: pde}
\begin{align}
\begin{split}
\partial_t^2 u
={}&
-\kappa^2
\Delta^2 u
-
2
\sigma_0
\partial_t
u
+
2
\sigma_1
\partial_t
\Delta
u
+{}
\\
&{}+
\kappa^2
\mathcal{L}(u, \phi)
+
\delta(\mathbf{r} - \mathbf{r}_{\mathrm{e}})
f_{\mathrm{e}}(t),
\end{split}
\label{eq: pde_disp}
\\
\Delta^2
\phi
={}&
-
\mathcal{L}(u, u)
.
\label{eq: pde_airy}
\end{align}
\end{subequations}
Here $u = u(\mathbf{r},t)$ is the transverse displacement and $\phi = \phi(\mathbf{r},t)$ is the Airy function. Both are defined for spatial coordinates $\mathbf{r} = (x,y) \in \mathcal{D} \subset \mathbb{R}^2$ and time $t \in \mathbb{R}^+$. Partial derivatives with respect to $x$, $y$ and $t$ are denoted as $\partial_x$, $\partial_y$ and $\partial_t$, respectively. The Laplacian operator is $\Delta = \partial_x^2 + \partial_y^2$ and the biharmonic operator is $\Delta^2$. The nonlinearity is introduced by the operator $\mathcal{L}$ defined for functions $\alpha(\mathbf{r})$ and $\beta(\mathbf{r})$ as:
\[
\mathcal{L}(\alpha, \beta)
=
\partial_x^2 \alpha \partial_y^2 \beta
+
\partial_y^2 \alpha \partial_x^2 \beta
-
2 \partial_x \partial_y \alpha \partial_x \partial_y \beta
.
\]

System~\eqref{eq: pde} is characterised by a stiffness parameter $\kappa$ and frequency-dependent loss parameters $\sigma_0 \geq 0$ and $\sigma_1 \geq 0$, all with dimensions of \unit{\sec\tothe{-1}}. It is excited at the location $\mathbf{r}_{\mathrm{e}} = (x_{\mathrm{e}}, y_{\mathrm{e}})$ by the function $f_{\mathrm{e}}(t)$ corresponding to a strike~\cite{Bilbao2023Gong}:
\[
f_{\mathrm{e}}(t)
=
\begin{cases}
\frac{1}{2}f_{\mathrm{amp}}
\big[
1
-
\cos
\big(
\frac{2 \pi t}{T_{\mathrm{e}}}
\big)
\big]
,
\quad
&t
\in
[
0,
T_{\mathrm{e}}
]
\\
0,
&\text{otherwise}
\end{cases}
\]
where $f_{\mathrm{amp}}$ is the excitation amplitude in \unit{\sec\tothe{-2}} and $T_{\mathrm{e}}$ is the excitation duration in \unit{\sec}.

We can obtain an energy balance for~\eqref{eq: pde_disp} by taking an $L^2(\mathcal{D})$ inner product with $\partial_t u$:
\[
\frac{d\mathfrak{H}}{dt}
=
\mathfrak{P}_{\mathrm{e}}
-
\mathfrak{P}_{\mathrm{d}}
+
\mathfrak{B}_{\mathrm{d}}
+
\mathfrak{B}_{\mathrm{lin}}
+
\mathfrak{B}_{\mathrm{nl}}
+
\mathfrak{B}_{\mathrm{nl}}'
.
\]

The energy $\mathfrak{H}$ is defined as:
\begin{equation}
\mathfrak{H}
=
\frac{1}{2}
\norm{\partial_t u}^2_{\mathcal{D}}
+
\frac{\kappa^2}{2}
\norm{\Delta u}^2_{\mathcal{D}}
+
\frac{\kappa^2}{4}
\norm{\Delta \phi}^2_{\mathcal{D}}
,
\label{eq: energy}
\end{equation}
where $\norm{\cdot}_{\mathcal{D}}$ is the $L^2(\mathcal{D})$ norm.

The terms $\mathfrak{P}_{\mathrm{e}}$ and $\mathfrak{P}_{\mathrm{d}}$ describe input and dissipated power, respectively, and are of the following form:
\begin{align*}
\mathfrak{P}_{\mathrm{e}}
&=
\partial_t
u(\mathbf{r}_{\mathrm{e}}, t)
f_{\mathrm{e}}(t)
,
\\
\mathfrak{P}_{\mathrm{d}}
&=
2
\sigma_0
\norm{\partial_t u}^2_{\mathcal{D}}
+
2
\sigma_1
\norm{\partial_t \boldsymbol{\nabla}_{\mathbf{r}} u}^2_{\mathcal{D}}
,
\end{align*}
where $\boldsymbol{\nabla}_{\mathbf{r}}$ is the gradient operator with respect to $\mathbf{r}$.

For a point on the boundary $\partial\mathcal{D}$ with normal and tangent unit vectors $\mathbf{n}$ and $\boldsymbol{\tau}$, respectively, we can write the boundary terms as~\cite{TorinPhD}:
\begingroup
\allowdisplaybreaks
\begin{subequations}\label{eq: boundary_terms}
\begin{align}
\mathfrak{B}_{\mathrm{d}}
={}&
2
\sigma_1
\oint_{\partial\mathcal{D}}
\partial_t 
u
\partial_n
\partial_t 
u
\;ds
,
\label{eq: boundary_diss}
\\
\mathfrak{B}_{\mathrm{lin}}
={}&
\kappa^2
\oint_{\partial\mathcal{D}}
(
\Delta
u
\partial_n
\partial_t
u
-
\partial_t 
u
\partial_n
\Delta
u
)
\;ds
,
\label{eq: boundary_lin}
\\
\mathfrak{B}_{\mathrm{nl}}
={}&
\frac{\kappa^2}{2}
\oint_{\partial\mathcal{D}}
(
\Delta
\partial_t
\phi
\partial_n \phi
-
\phi
\partial_n
\Delta
\partial_t
\phi
)
\;ds
\label{eq: boundary_nl_greens}
,
\\
\begin{split}
\mathfrak{B}_{\mathrm{nl}}'
={}&
\kappa^2
\oint_{\partial\mathcal{D}}
[
\partial_n
\partial_{\tau}
u
(
\phi
\partial_{\tau} \partial_t u
-
\partial_t u
\partial_{\tau} \phi
)
+{}\\
&{}+
(
\Delta u
-
\partial^2_n
u
)
(
\partial_t u
\partial_n \phi
-
\phi
\partial_n \partial_t u
)
]
\;ds
,
\end{split}
\label{eq: boundary_nl_adjointness}
\end{align}
\end{subequations}
\endgroup
where $\partial_n \triangleq \boldsymbol{\nabla}_{\mathbf{r}}\tran\mathbf{n}$ and $\partial_{\tau} \triangleq \boldsymbol{\nabla}_{\mathbf{r}}\tran\boldsymbol{\tau}$ are normal and tangential derivatives, respectively.

We consider a rectangular region $\mathcal{D} = [0,\sqrt{\eta}]\times[0,\sfrac{1}{\sqrt{\eta}}]$ with an aspect ratio $\eta > 0$ and employ the simply supported boundary conditions~\cite{Schaeffer1979}:
\begin{equation}
u
=
\Delta u
=
0,
\quad
\partial_n
\phi
=
\partial_n
\Delta \phi
=
0,
\quad
\mathbf{r}
\in
\partial\mathcal{D}.
\label{eq: boundary_cond}
\end{equation}
As can be seen, the boundary terms~\eqref{eq: boundary_terms} vanish under the boundary conditions~\eqref{eq: boundary_cond}. Note that the energy $\mathfrak{H}$ takes the form~\eqref{eq: energy} only under a fixed termination for the displacement $u$~\cite{BilbaoNSS}. Initial conditions are assumed to be zero, unless stated otherwise.

\section{Pseudospectral Method}\label{sec: ps}

\subsection{Modal Decomposition}

The boundary conditions~\eqref{eq: boundary_cond} allow a closed-form solution for modal shapes~\cite{Muradova2008}. We decompose the displacement $u$ and the Airy function $\phi$ using orthonormal sine and cosine series with modal coefficients $q_{ij}(t)$ and $\xi_{ij}(t)$, respectively:
\begin{subequations}\label{eq: decomps}
\begin{align}
u
&=
\sum_{i=1}^{M_{x\vphantom{y}}}
\sum_{j=1}^{M_y}
2
\sin(k_{x,i}x)
\sin(k_{y,j}y)
q_{ij}(t)
,
\label{eq: decomp_disp}
\\
\phi
&=
\sum_{i=0}^{M_{x\vphantom{y}}}
\sum_{j=0}^{M_y}
2
\alpha_{i}
\alpha_{j}
\cos(k_{x,i}x)
\cos(k_{y,j}y)
\xi_{ij}(t)
,
\label{eq: decomp_airy}
\end{align}
\end{subequations}
where $M_x$ and $M_y$ are the number of modes for each spatial direction. A scaling factor $\alpha_i$ is defined as $\alpha_0 = \sfrac{1}{\sqrt{2}}, \alpha_i = 1, i \neq 0$. Spatial components $k_{x,i}$ and $k_{y,j}$ of a wavevector and a wavenumber $k_{ij}$ are: 
\[
k_{x,i}
=
\frac{i \pi}{\sqrt{\eta}}
,
\quad
k_{y,j}
=
j \pi \sqrt{\eta}
,
\quad
k_{ij}
=
\sqrt{k_{x,i}^2 + k_{y,j}^2}
.
\]

Compared to the displacement $u$, the Airy function $\phi$ includes one-dimensional modes when either $i=0$ or $j=0$ and a constant mode when both $i=0$ and $j=0$. Without loss of generality, we zero out the constant mode $\xi_{00} \triangleq 0$ since the Airy function $\phi$ is defined up to an arbitrary bilinear function of spatial coordinates~\cite{Thomas2008}.

\subsection{Spatial Discretisation}

We consider a staggered spatial grid defined as:
\begin{equation}
x_i
=
\sqrt{\eta}
\frac{i + \frac{1}{2}}{N_x}
,
\quad
y_j
=
\frac{1}{{\sqrt{\eta}}}
\frac{j + \frac{1}{2}}{N_y}
,
\quad
\substack{
i=0,\dots,N_x-1\\
j=0,\dots,N_y-1
}
,
\label{eq: grid}
\end{equation}
where $N_x$ and $N_y$ are the number of collocation points for each spatial direction.
We define grid functions $u_{ij}(t)$ and $\phi_{ij}(t)$ that represent decompositions~\eqref{eq: decomps} evaluated on the grid~\eqref{eq: grid}. We can rewrite them in a vectorised form as:
\begin{align*}
\mathbf{u}
&=
\sqrt{N_xN_y}
(
\mathbf{S}_x
\otimes
\mathbf{S}_y
)\tran
\mathbf{q}
,
\\
\boldsymbol{\phi}
&=
\sqrt{N_xN_y}
(
\mathbf{C}_x
\otimes
\mathbf{C}_y
)\tran
\boldsymbol{\xi}
.
\end{align*}
Here vectors $\mathbf{u} \in \mathbb{R}^{N_x N_y}$, $\boldsymbol{\phi} \in \mathbb{R}^{N_x N_y}$, $\mathbf{q} \in \mathbb{R}^{M_x M_y}$ and $\boldsymbol{\xi} \in \mathbb{R}^{(M_x+1)(M_y+1)}$ are formed through row-major vectorisation of corresponding grid functions $u_{ij}(t)$ and $\phi_{ij}(t)$ and modal coefficients $q_{ij}(t)$ and $\xi_{ij}(t)$. Matrices $\mathbf{S}_a$ and $\mathbf{C}_a$ for $a \in \{x, y\}$ are truncated matrices of the orthonormal type-\rom{2} discrete sine and cosine transforms, respectively, defined as~\cite{BritanakDCTDST}:
\begin{alignat*}{2}
[\mathbf{S}_a]_{ij}
&=
\sqrt{
\frac{2}{N_a}
}
\sin
\bigg[
\frac{i (j + \frac{1}{2}) \pi}{N_a}
\bigg],
\quad
&&\substack{i=1,\dots,M_a\\j=0,\dots,N_a-1}
,
\\
[\mathbf{C}_a]_{ij}
&=
\sqrt{
\frac{2}{N_a}
}
\alpha_{i}
\cos
\bigg[
\frac{i (j + \frac{1}{2}) \pi}{N_a}
\bigg]
,
\quad
&&\substack{i=0,\dots,M_a\\j=0,\dots,N_a-1}
.
\end{alignat*}
Assuming $N_a > M_a$, rows of these matrices are orthonormal vectors so that $\mathbf{S}_a\mathbf{S}_a\tran = \mathbf{I}_{M_a}$ and $\mathbf{C}_a\mathbf{C}_a\tran = \mathbf{I}_{M_a + 1}$ where $\mathbf{I}_N$ denotes an $N \times N$ identity matrix. The Kronecker product~$\otimes$ is used to construct two-dimensional transforms.

\subsection{Modal Equations}

We substitute decompositions~\eqref{eq: decomps} into system~\eqref{eq: pde} and require that it is satisfied on the grid~\eqref{eq: grid}. Left-multiplying~\eqref{eq: pde_disp} and~\eqref{eq: pde_airy} after discretisation by $(\sfrac{1}{\sqrt{N_xN_y}})(\mathbf{S}_x\otimes\mathbf{S}_y)$ and $(\sfrac{1}{\sqrt{N_xN_y}})(\mathbf{C}_x\otimes\mathbf{C}_y)$, respectively, we obtain:
\begin{subequations}\label{eq: ode}
\begin{align}
&\ddot{\mathbf{q}}
+
2
\boldsymbol{\sigma}
\odot
\dot{\mathbf{q}}
+
\boldsymbol{\omega}^2
\odot
\mathbf{q}
=
\kappa^2
\mathbf{f}_1(\mathbf{q}, \boldsymbol{\xi})
+
\mathbf{j}_{\mathrm{e}}
f_{\mathrm{e}}(t)
,
\label{eq: ode_disp}
\\
&
\mathbf{k}^4
\odot
\boldsymbol{\xi}
=
-
\mathbf{f}_2(\mathbf{q}, \mathbf{q})
.
\label{eq: ode_airy}
\end{align}
\end{subequations}
Here a vector $\mathbf{k} \in \mathbb{R}^{(M_x + 1)(M_y + 1)}$ is a row-major vectorisation of wavenumbers $k_{ij},\;i=0,\dots,M_x,\;j=0,\dots,M_y$. In addition, we define $\hat{\mathbf{k}} \in \mathbb{R}^{M_x M_y}$ by excluding indices $i=0$ and $j=0$ before vectorisation. Modal frequencies $\boldsymbol{\omega} \in \mathbb{R}^{M_x M_y}$ and damping coefficients $\boldsymbol{\sigma} \in \mathbb{R}^{M_x M_y}$ are obtained as:
\[
\boldsymbol{\omega}
=
\kappa
\hat{\mathbf{k}}^2
,
\quad
\boldsymbol{\sigma}
=
\sigma_0
+
\sigma_1
\hat{\mathbf{k}}^2
.
\]
Note that damping coefficients $\boldsymbol{\sigma}$ can be generalised to represent an arbitrary loss profile. The Hadamard (elementwise) product~$\odot$ is used to multiply vectors and the power operator indicates elementwise raising. 

The function $f_{\mathrm{e}}(t)$ in~\eqref{eq: ode_disp} is weighted by modal shapes $2\sin(k_{x,i}x_{\mathrm{e}})\sin(k_{y,j}y_{\mathrm{e}})$ at the location $\mathbf{r}_{\mathrm{e}}$, consolidated into a vector $\mathbf{j}_{\mathrm{e}} \in \mathbb{R}^{M_x M_y}$. This vector represents coefficients of an orthonormal sine series for the delta function $\delta(\mathbf{r}-\mathbf{r}_{\mathrm{e}})$.

\subsection{Nonlinearity}

System~\eqref{eq: ode} is in the same form as in a modal approach~\cite{Ducceschi2015}. The difference lies in the definition of the nonlinear functions $\mathbf{f}_1(\mathbf{q}, \boldsymbol{\xi})$ and $\mathbf{f}_2(\mathbf{q}, \mathbf{q})$:
\begingroup
\allowdisplaybreaks
\begin{subequations}\label{eq: nl}
\begin{align}
\mathbf{f}_1(\mathbf{q}, \boldsymbol{\xi})
&=
\frac{1}{\sqrt{N_xN_y}}
(\mathbf{S}_x\otimes\mathbf{S}_y)
\mathbf{l}_1(\mathbf{u},\boldsymbol{\phi})
,
\label{eq: nl_1}
\\
\mathbf{f}_2(\mathbf{q}, \mathbf{q})
&=
\frac{1}{\sqrt{N_xN_y}}
(\mathbf{C}_x\otimes\mathbf{C}_y)
\mathbf{l}_2(\mathbf{u},\mathbf{u})
,
\label{eq: nl_2}
\end{align}
\end{subequations}
\endgroup
where discretised operators $\mathbf{l}_1(\mathbf{u},\boldsymbol{\phi})$ and $\mathbf{l}_2(\mathbf{u},\mathbf{u})$ give exact values of the operator $\mathcal{L}$ on the grid~\eqref{eq: grid}. To define these discretised operators, we require differentiation matrices for grid functions $\mathbf{u}$ and $\boldsymbol{\phi}$ that exactly evaluate derivatives of the displacement $u$ and the Airy function $\phi$ on the grid~\eqref{eq: grid}, given decompositions~\eqref{eq: decomps}.

First, we obtain differentiation matrices $\mathbf{D}_{xx,s}$ and $\mathbf{D}_{xx,c}$ for derivatives $\partial_x^2u$ and $\partial_x^2\phi$, respectively:
\begin{align*}
\mathbf{D}_{xx,s}
&=
-
\big(
\mathbf{S}_x\tran
\hat{\mathbf{K}}_x^2
\mathbf{S}_x
\otimes
\mathbf{I}_{N_y}
\big)
,
\\
\mathbf{D}_{xx,c}
&=
-
\big(
\mathbf{C}_x\tran
\mathbf{K}_x^2
\mathbf{C}_x
\otimes
\mathbf{I}_{N_y}
\big)
,
\end{align*}
where $\mathbf{K}_x = \diag(k_{x,0},\dots,k_{x,M_x})$ and $\hat{\mathbf{K}}_x$ is the matrix $\mathbf{K}_x$ without the wavenumber $k_{x,0}$. Note that we can split $\mathbf{D}_{xx,s} = \mathbf{D}_{x,\hat{c}}\mathbf{D}_{x,s}$ into first-order differentiation matrices defined as:
\[
\mathbf{D}_{x,s}
=
\big(
\hat{\mathbf{C}}_x\tran
\hat{\mathbf{K}}_x
\mathbf{S}_x
\otimes
\mathbf{I}_{N_y}
\big),
\quad
\mathbf{D}_{x,\hat{c}}
=
-
\mathbf{D}_{x,s}
\tran
,
\]
where $\hat{\mathbf{C}}_x$ is the matrix $\mathbf{C}_x$ without the first row. This is not be possible for $\mathbf{D}_{xx,c}$ since it accounts for one-dimensional modes and a constant mode which are not present in a discrete sine transform. By analogy, we can construct differentiation matrices $\mathbf{D}_{yy,s}$, $\mathbf{D}_{yy,c}$, $\mathbf{D}_{y,s}$ and $\mathbf{D}_{y,\hat{c}}$ in the $y$-direction.

Second, we obtain differentiation matrices $\mathbf{D}_{xy,s}$ and $\mathbf{D}_{xy,\hat{c}}$ for derivatives $\partial_x\partial_y u$ and $\partial_x\partial_y \phi$, respectively:
\[
\mathbf{D}_{xy,s}
=
\mathbf{D}_{x,s}
\mathbf{D}_{y,s}
,
\quad
\mathbf{D}_{xy,\hat{c}}
=
\mathbf{D}_{xy,s}
\tran
,
\]
where $\mathbf{D}_{xy,\hat{c}}\boldsymbol{\phi}$ is exact since one-dimensional modes and a constant mode do not affect the value of $\partial_x\partial_y \phi$.

Finally, we can define $\mathbf{l}_1(\mathbf{u},\boldsymbol{\phi})$ and $\mathbf{l}_2(\mathbf{u},\mathbf{u})$ as follows:
\begin{align*}
\begin{split}
\mathbf{l}_1(\mathbf{u}, \boldsymbol{\phi})
={}&
\mathbf{D}_{xx,s}
\mathbf{u}
\odot
\mathbf{D}_{yy,c}
\boldsymbol{\phi}
+
\mathbf{D}_{yy,s}
\mathbf{u}
\odot
\mathbf{D}_{xx,c}
\boldsymbol{\phi}
-{}
\\
&{}-
2
\mathbf{D}_{xy,s}
\mathbf{u}
\odot
\mathbf{D}_{xy,\hat{c}}
\boldsymbol{\phi}
,
\end{split}
\\
\begin{split}
\mathbf{l}_2(\mathbf{u}, \mathbf{u})
={}&
\mathbf{D}_{xx,s}
\mathbf{u}
\odot
\mathbf{D}_{yy,s}
\mathbf{u}
+
\mathbf{D}_{yy,s}
\mathbf{u}
\odot
\mathbf{D}_{xx,s}
\mathbf{u}
-{}
\\
&{}-
2
\mathbf{D}_{xy,s}
\mathbf{u}
\odot
\mathbf{D}_{xy,s}
\mathbf{u}
.
\end{split}
\end{align*}

To mitigate aliasing errors due to products in the spatial domain, we employ the $\sfrac{3}{2}$ rule~\cite{BoydSpectral} by setting $N_a = \ceil{\sfrac{3M_a}{2}} + 1,\; a \in \{x, y\}$. Under this choice aliasing would occur only for the highest third of wavenumbers which are filtered out by discrete sine and cosine transforms in~\eqref{eq: nl}. For the next section, it is useful to define ``brick-wall'' filters in the spatial domain as symmetric matrices $\mathbf{H}_s \triangleq (\mathbf{S}_x\otimes\mathbf{S}_y)\tran(\mathbf{S}_x\otimes\mathbf{S}_y)$ and $\mathbf{H}_c \triangleq (\mathbf{C}_x\otimes\mathbf{C}_y)\tran(\mathbf{C}_x\otimes\mathbf{C}_y)$.

The computational cost of nonlinearity is mainly determined by discrete sine and cosine transforms. For simplicity, assume a square plate so that number of modes and collocation points is the same in both spatial directions. Applying transforms independently in each dimension using naive matrix multiplication, it takes $\mathcal{O}(N_x^3)$ operations to evaluate $\mathbf{f}_1(\mathbf{q}, \boldsymbol{\xi})$. This estimate can be reduced to $\mathcal{O}(N_x^2 \log_2 N_x)$ if fast transforms are employed~\cite{BritanakDCTDST}. In contrast, it can be shown that a modal approach~\cite{Ducceschi2015} would require evaluation of $\mathcal{O}(M_x^4) = \mathcal{O}(N_x^4)$ coupling terms with products $q_{ij}\xi_{lm}$. This indicates that the pseudospectral method is significantly more efficient for a large number of modes.

\subsection{Nonlinear Potential Energy}

In view of using a scalar auxiliary variable technique, we require a non-negative potential $V(\mathbf{q})$ so that:
\[
\mathbf{f}_1(\mathbf{q}, \boldsymbol{\xi})
=
-\boldsymbol{\nabla}_{\mathbf{q}}V(\mathbf{q})
.
\]
To derive this potential, we need to specify product rules for differentiation matrices.

Assume grid functions $\boldsymbol{\alpha}$ and $\boldsymbol{\beta}$ that are bandlimited with respect to a transform $(\mathbf{S}_x\otimes\mathbf{S}_y)$ and a grid function $\boldsymbol{\gamma}$ that is bandlimited with respect to a transform $(\hat{\mathbf{C}}_x\otimes\hat{\mathbf{C}}_y)$. Considering product-to-sum identities for sine and cosine functions, the following product rules hold:
\begin{align*}
\begin{split}
\mathbf{D}_{xx,c}
\mathbf{H}_{c}
(
\boldsymbol{\alpha}
\odot
\boldsymbol{\beta}
)
&=
\mathbf{H}_{c}
(
\mathbf{D}_{xx,s}
\boldsymbol{\alpha}
\odot
\boldsymbol{\beta}
+
\boldsymbol{\alpha}
\odot
\mathbf{D}_{xx,s}
\boldsymbol{\beta}
+{}
\\
{}+{}&
2
\mathbf{D}_{x,s}
\boldsymbol{\alpha}
\odot
\mathbf{D}_{x,s}
\boldsymbol{\beta}
)
,
\end{split}
\\
\mathbf{D}_{xy,s}
\mathbf{H}_{s}
(
\boldsymbol{\alpha}
\odot
\boldsymbol{\gamma}
)
&{}={}
\mathbf{H}_{c}
(
\mathbf{D}_{xy,s}
\boldsymbol{\alpha}
\odot
\boldsymbol{\gamma}
+
\boldsymbol{\alpha}
\odot
\mathbf{D}_{xy,\hat{c}}
\boldsymbol{\gamma}
+{}
\\
{}+{}&
\mathbf{D}_{x,s}
\boldsymbol{\alpha}
\odot
\mathbf{D}_{y,\hat{c}}
\boldsymbol{\gamma}
+
\mathbf{D}_{y,s}
\boldsymbol{\alpha}
\odot
\mathbf{D}_{x,\hat{c}}
\boldsymbol{\gamma}
)
.
\end{align*}
By analogy, we can define a product rule for $\mathbf{D}_{yy,c}$.

Taking a dot product of $\mathbf{f}_1(\mathbf{q}, \boldsymbol{\xi})$ with $\dot{\mathbf{q}}$, exploiting the fact that ``brick-wall'' filters do not change bandlimited grid functions and using the product rules, we obtain: 
\begin{multline*}
\dot{\mathbf{q}}\tran
\mathbf{f}_1(\mathbf{q}, \boldsymbol{\xi})
=
\frac{1}{N_xN_y}
\dot{\mathbf{u}}\tran
\mathbf{l}_1(\mathbf{u}, \boldsymbol{\phi})
=
\\
\begin{aligned}
{}={}&
\frac{1}{N_xN_y}
(
\mathbf{H}_c
\mathbf{D}_{yy,c}
\boldsymbol{\phi}
)\tran
(
\mathbf{D}_{xx,s}
\mathbf{u}
\odot
\dot{\mathbf{u}}
)
+{}
\\
{}+{}&
\frac{1}{N_xN_y}
(
\mathbf{H}_c
\mathbf{D}_{xx,c}
\boldsymbol{\phi}
)\tran
(
\mathbf{D}_{yy,s}
\mathbf{u}
\odot
\dot{\mathbf{u}}
)
-{}
\\
{}-{}&
\frac{2}{N_xN_y}
(
\mathbf{H}_s
\mathbf{D}_{xy,\hat{c}}
\boldsymbol{\phi}
)\tran
(
\mathbf{D}_{xy,s}
\mathbf{u}
\odot
\dot{\mathbf{u}}
)
={}
\end{aligned}
\\
{}=
\frac{1}{N_xN_y}
\boldsymbol{\phi}\tran
\mathbf{H}_c
\mathbf{l}_2(\mathbf{u}, \dot{\mathbf{u}})
=
\boldsymbol{\xi}\tran
\mathbf{f}_2(\mathbf{q}, \dot{\mathbf{q}})
,
\end{multline*}
which represents a discretised ``triple self-adjointness'' property of the operator $\mathcal{L}$~\cite{Thomas2008}. Using the symmetry property $\mathbf{f}_2(\mathbf{q}, \dot{\mathbf{q}}) = \mathbf{f}_2(\dot{\mathbf{q}},\mathbf{q})$, we obtain:
\begin{multline*}
\boldsymbol{\xi}\tran
\mathbf{f}_2(\mathbf{q}, \dot{\mathbf{q}})
=
\boldsymbol{\xi}\tran
\frac{1}{2}
\frac{d}{dt}
\mathbf{f}_2(\mathbf{q}, \mathbf{q})
={}\\
\overset{\eqref{eq: ode_airy}}{=}
\boldsymbol{\xi}\tran
\frac{1}{2}
\frac{d}{dt}
(
-\mathbf{k}^4
\odot
\boldsymbol{\xi}
)
=
-
\dot{V}(\mathbf{q})
.
\end{multline*}
The potential $V(\mathbf{q})$ is defined as:
\[
V(\mathbf{q})
=
\frac{1}{4}
\norm{
\mathbf{k}^2
\odot
\boldsymbol{\xi}
(\mathbf{q})
}_2^2
\geq
0
,
\]
where $\norm{\cdot}_p$ for $p\geq 1$ is the $p$-norm.
To evaluate $V(\mathbf{q})$, we substitute the solution for $\boldsymbol{\xi} = \boldsymbol{\xi}(\mathbf{q})$ from~\eqref{eq: ode_airy}, where the biharmonic operator can be easily inverted by elementwise division for all but the constant mode. It can be shown that $\mathbf{f}_2(\mathbf{q}, \mathbf{q})$ has a zero constant mode so that~\eqref{eq: ode_airy} is fully satisfied.

\section{Numerical Solver}\label{sec: solver}

\subsection{Quadratisation}

We introduce an auxiliary variable \mybox{$\psi \triangleq \sqrt{2V(\mathbf{q}) + \epsilon}$} for an arbitrary constant $\epsilon > 0$ and rewrite~\eqref{eq: ode_disp} as:
\begin{subequations}\label{eq: ode_quad}
\begin{align}
&\ddot{\mathbf{q}}
+
2
\boldsymbol{\sigma}
\odot
\dot{\mathbf{q}}
+
\boldsymbol{\omega}^2
\odot
\mathbf{q}
=
-
\kappa^2
\psi
\mathbf{g}
+
\mathbf{j}_{\mathrm{e}}
f_{\mathrm{e}}(t)
,
\label{eq: ode_quad_disp}
\\
&
\dot{\psi}
=
\mathbf{g}\tran
\dot{\mathbf{q}}
,
\label{eq: ode_quad_aux}
\end{align}
\end{subequations}
where $\mathbf{g} \triangleq \mathbf{g}_{\mathrm{std}}(\mathbf{q}) = \nabla_{\mathbf{q}}\psi$. Furthermore, we add a control term to~\eqref{eq: ode_quad} as described by Risse~et~al.\@~\cite{Risse2025} to reduce the drift between numerical values of $\psi$ and $\sqrt{2 V(\mathbf{q}) + \epsilon}$ in discrete time. This results in modification of the coupling term $\mathbf{g}$:
\begin{align*}
&\mathbf{g}
\triangleq
\mathbf{g}_{\mathrm{std}}(\mathbf{q})
+
\mathbf{g}_{\mathrm{mod}}(\mathbf{q}, \dot{\mathbf{q}}, \psi)
,
\\
&\mathbf{g}_{\mathrm{mod}}(\mathbf{q}, \dot{\mathbf{q}}, \psi)
=
-\lambda_0
\Big(
\psi
-
\sqrt{2 V(\mathbf{q}) + \epsilon}
\Big)
\frac{
\sign(\dot{\mathbf{q}})
}{
\norm{\dot{\mathbf{q}}}_1
}
,
\end{align*}
where $\lambda_0 \geq 0$ is the control parameter in \unit{\sec\tothe{-1}}.

\subsection{Time Discretisation}

We choose a time step $T$ in \unit{\sec}, yelding a sampling rate $f_{\mathrm{s}} = \sfrac{1}{T}$. Then, we approximate $\mathbf{q}(t)$ and $\psi(t)$ by time series $\mathbf{q}^n$ and $\psi^{n-\sfrac{1}{2}}$ on interleaved temporal grids $t^n = nT$ and $t^{n-\sfrac{1}{2}} = (n - \sfrac{1}{2})T$ for $n \in \mathbb{N}$. We introduce one-step difference and averaging operators $\delta_{t\pm}$ and $\mu_{t\pm}$, acting on time series as:
\[
\delta_{t\pm}
\mathbf{q}^n
=
\pm
\frac{\mathbf{q}^{n\pm 1} - \mathbf{q}^n}{T}
,
\quad
\mu_{t\pm}
\mathbf{q}^n
=
\frac{\mathbf{q}^{n\pm 1} + \mathbf{q}^n}{2}
.
\]
Using centred approximations $\delta_{t\cdot} = \delta_{t+}\mu_{t-}$ and $\delta_{tt} = \delta_{t+}\delta_{t-}$ for first-order and second-order time derivatives, respectively, we define the scheme as:
\begin{subequations}\label{eq: ode_quad_discr}
\begin{align}
\begin{split}
&\delta_{tt}\mathbf{q}^n
+
2
\tilde{\boldsymbol{\sigma}}
\odot
\delta_{t\cdot}
\mathbf{q}^n
+
\tilde{\boldsymbol{\omega}}^2
\odot
\mathbf{q}^n
={}
\\
&
\hphantom{\delta_{t+}\psi^{n-\sfrac{1}{2}}}
{}=
-
\kappa^2
\mu_{t+}
\psi^{n-\sfrac{1}{2}}
\mathbf{g}^n
+
\mathbf{j}_{\mathrm{e}}
f_{\mathrm{e}}^n
,
\end{split}
\label{eq: ode_quad_discr_disp}
\\
&
\delta_{t+}
\psi^{n-\sfrac{1}{2}}
=
(\mathbf{g}^n)\tran
\delta_{t\cdot}
\mathbf{q}^n
\label{eq: ode_quad_discr_aux}
,
\end{align}
\end{subequations}
where
$
\mathbf{g}^n
=
\mathbf{g}_{\mathrm{std}}
(\mathbf{q}^n)
+
\mathbf{g}_{\mathrm{mod}}
(\mu_{t-}\mathbf{q}^n, \delta_{t-}\mathbf{q}^n, \psi^{n-\sfrac{1}{2}})$
and
$f_{\mathrm{e}}^n = f_{\mathrm{e}}(t^n)$. 
In addition, modal frequencies and damping coefficients are adjusted to mitigate numerical dispersion and ensure exact discretisation of the linear part of~\eqref{eq: ode_quad_disp}~\cite{BilbaoNSS}:
\begin{align*}
\tilde{\omega}_{ij}^2
&=
\frac{2}{T^2}
\frac{
1
-
2
e^{-\sigma_{ij} T}
\cos
\Big(
T
\sqrt{
\smash[b]{
\omega_{ij}^2
-
\sigma_{ij}^2
}
}
\Big)
+
e^{-2\sigma_{ij} T}
}{
1
+
e^{-2\sigma_{ij} T}
}
,
\\
\tilde{\sigma}_{ij}
&=
\frac{1}{T}
\frac{
1
-
e^{-2\sigma_{ij} T}
}{
1
+
e^{-2\sigma_{ij} T}
}
.
\end{align*}

We can obtain an energy balance for~\eqref{eq: ode_quad_discr_disp} by taking a dot product with $\delta_{t\cdot}\mathbf{q}^n$ and employing product identities for discrete-time operators~\cite{BilbaoNSS}:
\[
\delta_{t+}
H^{n-\sfrac{1}{2}}
=
\underbrace{
(\delta_{t\cdot}\mathbf{q}^n)\tran
\mathbf{j}_{\mathrm{e}}
f_{\mathrm{e}}^n
}_{\triangleq P_{\mathrm{e}}^n}
-
\underbrace{
(\delta_{t\cdot}\mathbf{q}^n)\tran
(
2\tilde{\boldsymbol{\sigma}}
\odot
\delta_{t\cdot}\mathbf{q}^n
)
}_{\triangleq P_{\mathrm{d}}^n}
,
\]
where the energy $H^{n-\sfrac{1}{2}}$ is defined as:
\begin{multline*}
H^{n-\sfrac{1}{2}}
=
\frac{1}{2}
\norm{
\delta_{t-}
\mathbf{q}^n
}_2^2
+
\frac{1}{2}
(\mathbf{q}^n)\tran
(
\tilde{\boldsymbol{\omega}}^2
\odot
\mathbf{q}^{n-1}
)
+{}\\
{}+
\frac{\kappa^2}{2}
\big(
\psi^{n-\sfrac{1}{2}}
\big)^{2}
.
\end{multline*}
The second term in $H^{n-\sfrac{1}{2}}$ can be bounded to guarantee non-negativity of numerical energy~\cite{Bilbao2023SAV}, resulting in a stability condition $\omega_{ij} < \pi f_{\mathrm{s}}$, i.e, frequencies up to a Nyquist limit can be reproduced. Given the definition of $\omega_{ij}$, the number of modes in simulation would then be smaller than the product $M_xM_y$.

To initialise the scheme, we require values $\mathbf{q}^0$ and $\mathbf{q}^1$ and set the auxiliary variable as $\psi^{\sfrac{1}{2}} = \sqrt{2 V(\mu_{t-}\mathbf{q}^1) + \epsilon}$. For time stepping, \eqref{eq: ode_quad_discr} can be written in an explicit update form after a matrix inversion using the Sherman-Morrison formula~\cite{Russo2025}.

\section{Numerical Results}\label{sec: results}

\begin{figure}[t]
\centering
\includegraphics[width=\columnwidth]{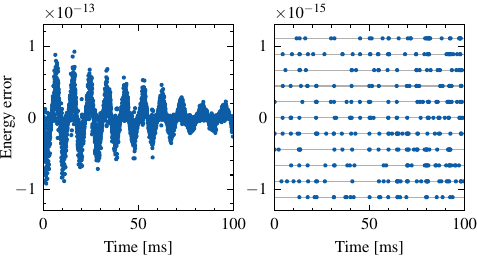}
\caption{Relative energy error $E^n$ (left) and its enlarged section that reveals numerical quantisation (right). Horizontal lines indicate multiples of machine accuracy $2^{-52}\approx \num{2.2e-16}$.}
\label{fig: energy}
\end{figure}

The source code used for numerical tests can be found in the accompanying repository\footnote{\url{https://github.com/victorzheleznov/fa2026}}. All simulations were done at $f_{\mathrm{s}} = \qty{44.1}{\kilo\hertz}$. A rectangular plate was used with a ratio $\eta = 1.1$. Frequency-dependent loss parameters were set as $\sigma_0 = 1.3$ and $\sigma_1 = \num{1e-4}$, resulting in a $T_{60}$ decay time of around $\qty{10}{\sec}$ for lower frequencies. The simulated frequency range was reduced to $\qty{17}{\kilo\hertz}$ away from the Nyquist limit to avoid anomalous behaviour of the scheme~\cite{BilbaoNSS}.

As a first case study, a large plate with $\kappa = 8$ was used, resulting in $1027$ modes. To check energy conservation, the plate was initialised in its first mode of vibration as $q_{00} = 5$ without additional excitation, i.e., $P_{\mathrm{e}}^n \equiv 0$. Step-by-step relative energy error $E^n$ is defined as:
\[
E^n
=
\frac{
H^{n+\sfrac{1}{2}}_{\mathrm{tot}}
-
H^{n-\sfrac{1}{2}}_{\mathrm{tot}}
}{
\floor{
H^{n-\sfrac{1}{2}}_{\mathrm{tot}}
}_2
},
\;
H^{n-\sfrac{1}{2}}_{\mathrm{tot}}
=
H^{n-\sfrac{1}{2}}
+
\sum_{i=1}^{n-1}
T
P_{\mathrm{d}}^{i}
,
\]
where $\floor{\cdot}_2$ denotes the nearest power of two rounded towards zero. As demonstrated on Fig.\@~\ref{fig: energy}, the energy error remains constant to machine accuracy during the simulation.

\begin{figure}[t]
\centering
\includegraphics[width=\columnwidth]{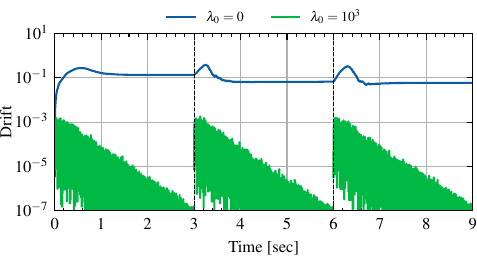}
\caption{Relative drift $D^{n-\sfrac{1}{2}}$. Vertical dashed lines indicate the moments in time when the plate is excited.}
\label{fig: drift}
\end{figure}

To test drift regulation, the plate was repeatedly excited using the function $f_{\mathrm{e}}(t)$ with $f_{\mathrm{amp}} = \num{2e6}$ and $T_{\mathrm{e}} = \qty{2}{\milli\s}$ at the location $\mathbf{r}_{\mathrm{e}} = (\num{0.17}, \num{0.42})$. The relative drift $D^{n-\sfrac{1}{2}}$ is defined as:
\[
D^{n-\sfrac{1}{2}}
=
\frac{
\psi^{n-\sfrac{1}{2}}
-
\sqrt{2 V(\mu_{t-}\mathbf{q}^n) + \epsilon}
}{
\max\limits_n
\sqrt{2 V(\mu_{t-}\mathbf{q}^n) + \epsilon}
}
,
\]
where the maximum value is taken over the whole duration of simulation. Comparison between the cases with and without drift regulation is given in Fig.\@~\ref{fig: drift}. As can be seen, the maximum value of the relative drift is reduced by two orders of magnitude and the drift steadily decreases towards zero by the control term with $\lambda_0 = \num{1e3}$ up to the next excitation. In contrast, the auxiliary variable remains incorrect over the whole duration of simulation in the case $\lambda_0=0$.

\begin{figure*}[t]
\centering
\includegraphics[width=\textwidth]{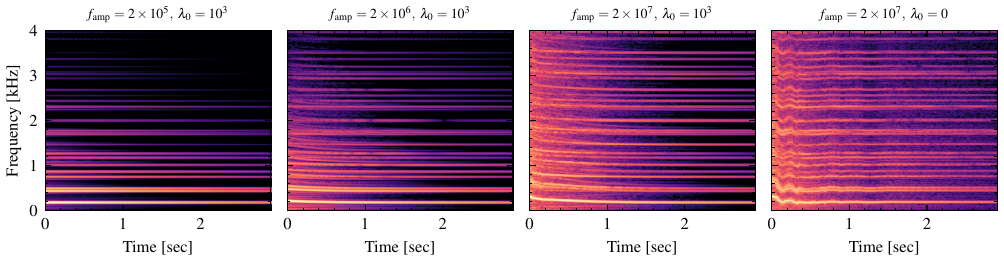}
\caption{Spectrograms for displacement of the plate at increasing excitation amplitudes, as indicated. The third and fourth spectrograms are given at identical excitation amplitudes to demonstrate drift regulation of the numerical solver.}
\label{fig: spec}
\end{figure*}

As a second case study, a small plate with $\kappa = 60$ was considered, resulting in $127$ modes. Fig.\@~\ref{fig: spec} provides spectrograms for displacement of the plate at increasing excitation amplitudes, taken at the location $\mathbf{r}_{\mathrm{o}} = (\num{0.64}, \num{0.79})$. Under a low amplitude, linear behaviour is recovered with distinct modal frequencies. At a higher amplitude, pitch glide effects appear. Then, a wideband noise is observed, resulting in a crash-like sound. As demonstrated on the rightmost spectrogram in Fig.\@~\ref{fig: spec}, instantaneous frequencies are significantly affected by the removal of the control term, leading to audible artefacts. Readers are encouraged to listen to sound examples presented on the accompanying page\footnote{\url{https://victorzheleznov.github.io/fa2026}}.

\section{Conclusion}

An explicit and stable pseudospectral method for simulation of the \FvK{} equations has been presented here. Numerical tests at an audio rate have confirmed energy conservation of the scheme and drift regulation properties of the scalar auxiliary variable technique, which were previously reported only for the case of nonlinear string vibration~\cite{Risse2025}. Additional work is required to evaluate real-time capability of the proposed method in comparison to a finite difference approach~\cite{Bilbao2023Gong} using an optimised C\texttt{++} implementation. In addition, extension to other boundary conditions is needed to further evaluate suitability of pseudospectral methods to problems in musical acoustics.

\section{Acknowledgments}

This work was supported by the SGSAH AHRC Doctoral Training Partnership [grant number AH/R012717/1]; and the University of Edinburgh. The authors would like to thank Olivier Thomas for discussions on boundary conditions for the \FvK{} equations.

\printbibliography

\end{document}